\documentstyle[sprocl]{article}
\input{psfig}
\def\Journal#1#2#3#4{{#1} {\bf #2}, #3 (#4)}

\def\PLB{{\em Phys. Lett.}  B}
\def\PRL{\em Phys. Rev. Lett.}
\def\PRD{{\em Phys. Rev.} D}
\def\ZPC{{\em Z. Phys.} C}

\def\be{\begin{equation}}
\def\ee{\end{equation}}
\def\bea{\begin{eqnarray}}
\def\eea{\end{eqnarray}}

\renewcommand{\textfraction}{0.1}
\renewcommand{\topfraction}{2.5}

\begin{document}
\renewcommand{\textfraction}{0.1}
\renewcommand{\topfraction}{2.5}

\title{ SQUARKS IN TEVATRON DILEPTON EVENTS ?}

\author{R. MICHAEL BARNETT$^1$\footnote{presented by RMB at the
DPF-96 Meeting, Minneapolis, Aug. 10-15}, 
LAWRENCE J. HALL$^{1,2}$}

\address{$^1$
     Lawrence Berkeley National Laboratory,
     University of California, Berkeley, California 94720\\
$^2$ Department of Physics,
     University of California, Berkeley, California 94720}

\maketitle\abstracts{
We consider unusual events in the CDF and D0 
dilepton+jets sample with very high 
$E_T$(lepton) and $E_T$(missing). 
It is possible, but very unlikely, that these events
originate from top quark pair production; however, they have
characteristics that are better accounted for by decays of
supersymmetric quarks with mass in the region of 300 GeV.}

For ten years, two primary signatures for squarks and gluinos have been
discussed\cite{THEORY,SS}.  The first consists of events with jets plus large
missing transverse energy.  The second consists of
multilepton events with jets and missing transverse energy 
($\!\not\!\!E_T$).  We have looked at the events reported by the CDF
and D0 collaborations in their top quark to dileptons
sample\cite{top}, and
find that two CDF events and one D0 event have characteristics
significantly different from other events in the top quark sample 
and similar to the second class of signature events for squarks and
gluinos.
While we cannot rule out a  
statistical fluctuation of the top signal or detector effects, we
feel it worthwhile to examine the consequences for supersymmetry if
these events are the precursors of a real supersymmetry signal.
This work was first reported in 
Ref.~4; more details can be found there.

In 110 pb$^{-1}$, the CDF Collaboration observed 10 such opposite-sign dilepton
events, where 6 were expected from $\bar{t} t$ production with $m_t = 175$ GeV,
and 2 events were expected from non-top Standard Model backgrounds
\cite{conf,TARTARELLI,Kruse}. 
The D0 collaboration has also
observed dilepton events, one of which
appears to have similar or even more dramatic characteristics \cite{Cochran}.
However, there is a large uncertainty in the 
measurement of the muon $E_T$ and of $\!\not\!\!E_T$.

These events which we label A, B, and C (C being the D0 event) have
large values of $E_S \equiv E_T^{\ell_1} + E_T^{\ell_2} +
\!\not\!\!E_T$, see Fig.~1a.
Event A  contains a third isolated 
charged track, which is likely to be an
electron. A third isolated, hard charged lepton would make this event
inconsistent with a $\bar{t} t$ origin. 
A kinematic argument 
shows that the values of  $E_T^{\ell_1}$, $E_T^{\ell_2}$ 
and $\not\!\!E_T$ of event B 
cannot arise from the decay of any pair of $W$'s whether or not 
the $W$'s originated in 
$\bar{t} t $ production (neglecting neutrinos in the jets).
This is also evident in Fig. 2 which shows $\!\not\!\!E_T$ with cuts
satisfied by event B:
$E_T^{\ell_1}\cos\theta(\ell_1-\!\not\!\!E_T) > 100$ GeV and 
$E_T^{\ell_2}\cos\theta(\ell_2-\!\not\!\!E_T) > 40$ GeV. 

In Figs. 1b-d we show three additional plots that demonstrate that a
top quark explanation for these events is quite unlikely.  In Fig.
1b we require large $E_S$ and see that top quark events are unlikely
to yield a small transverse angle between the leptons.  Figs. 1c and
1d make no cuts and show very good fits to most of the reported CDF
top quark events, but events A-C are on the tails of the
distributions.

Heavier squarks and gluinos often decay via a sequence of
cascades through charginos $(\widetilde{\chi}^+)$ and neutralinos  
$(\widetilde{\chi}^0)$, yielding events with isolated charged leptons $\ell$ 
as well as jets and $\not\!\!E_T$ \cite{THEORY,SS}. The isolated charged leptons
can arise from both $\widetilde{\chi}^+,\widetilde{\chi}^0$ decays, such as
$\widetilde{\chi}^+ \rightarrow \widetilde{\nu} \bar{e}$,
$\widetilde{e} \bar{\nu}$ and $\widetilde{\chi}^0 \rightarrow \widetilde{e} 
\bar{e}$, $\widetilde{\nu} \bar{\nu}$, and also from slepton decays, for 
example $\widetilde e\rightarrow e \widetilde\chi_1^0.$
There are plausible ranges of superpartner masses in which
the cascade decays of squarks, $\widetilde{q}
\rightarrow  \widetilde{\chi} \rightarrow  \widetilde{\ell} \rightarrow \ell$, 
could lead to a few  $\ell \ell (\ell) jj \!\not\!\!E_T$ events,
with extraordinarily high  $\not\!\!E_T$ and 
$E_T^\ell$, in the last Tevatron run.


\begin{figure} 
\psfig{figure=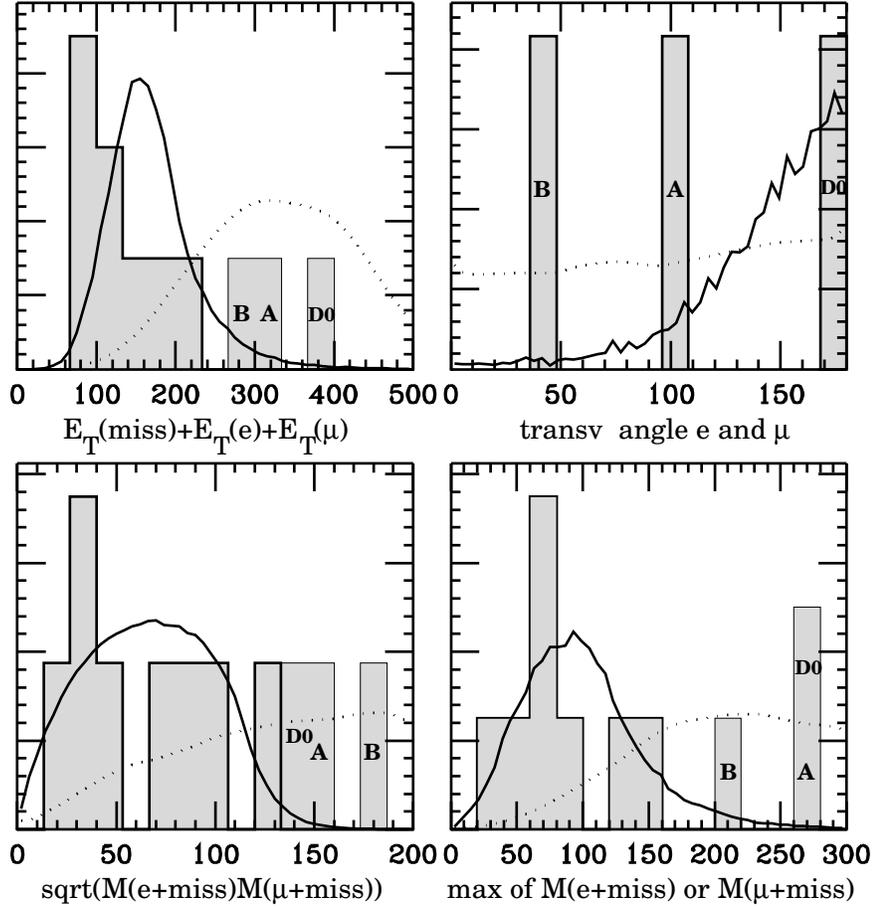,width=4.5in} 
\vskip -0.8in
\caption{
Expected distributions for (a) $E_S= 
E_T^{\ell_1}+E_T^{\ell_2}+ \not\!\!E_T$ , (b) 
$\theta_T$ between the two leptons for $E_S> 250$ GeV, (c) the
product of the transverse masses of $\ell_1 + \not\!\!E_T$ and
$\ell_2 + \not\!\!E_T$, and (d) the maximum of the two transverse
masses in c).  The solid curves are for
$t\bar t$  production.  The dotted curve
has both leptons from $\widetilde \ell \rightarrow \ell 
\widetilde\chi_1^0$ decays.  The histograms show the CDF data.
The three events mentioned in the text
are labelled A, B, and D0.  
\label{fig:dists}}
\end{figure}

%

We study the simplified case where the three 
$\widetilde{\chi}'$ states ($\widetilde\chi$ states relevant to
$\widetilde q$ decays) are dominantly the $SU(2)_L$ gauginos: 
$\widetilde{\chi}_1^{\pm} \approx \widetilde{w}^{\pm}$ and  
$\widetilde{\chi}_2^0 \approx \widetilde{w}_3$, choosing 
$|\mu| > 300$ GeV (for $M_2 = 260$ GeV).

In this scenario there are five flavors of left-handed squarks with masses in 
the region of 310 GeV. These decay to $SU(2)_L$ gauginos,
$\widetilde{\chi}_1^{\pm}$ and $\widetilde{\chi}^0_2$, 
of mass near 260 GeV, which in turn decay to left-handed 
sleptons with mass near 220 GeV. The hardest charged leptons are
produced in the 
final cascade of the sleptons to the LSP $\widetilde{\chi}^0_1$,
taken to be dominantly bino..
The $\widetilde{\chi}_1^0$ mass is given by the hypercharge gaugino
mass parameter, $M_1$, which is therefore several times less than
$M_2$. The region of parameters of interest to us does not allow the
relation $M_2 \approx 2 M_1$, which occurs in simple schemes of
grand unification with large messenger scales for supersymmetry
breaking.

In our scheme, dilepton events, such as events $A$ and $B$, arise
from the decay of
a $\widetilde{q}_L^{(\dagger)} \widetilde{q}_L$ pair. 
Since $\widetilde{g}$ decays
to $ q^\dagger \widetilde{q}$ or to $q \widetilde{q}^\dagger$, events from
$\widetilde{g} \widetilde{q}$ and $\widetilde{g} \widetilde{q}^\dagger$
production  look similar to $\widetilde{q}^\dagger \widetilde{q}$
events.
Because $\widetilde{q}_L$ has a small hypercharge, the direct decay 
$\widetilde{q}_L \rightarrow q \widetilde{\chi}^0_1$ has a small branching
ratio compared to the cascade mode $\widetilde{q}_L \rightarrow 
\widetilde{\chi}^0_2, \widetilde{\chi}^+_1 \rightarrow \widetilde{\ell}
\rightarrow \ell \widetilde{\chi}^0_1$, 
These events have the number of
isolated charged leptons, $N_L$, varying from 0 to 4.
Neglecting phase space, $3\over 4$ 
of these $\widetilde{q}_L^{(\dagger)} \widetilde{q}_L$ events have
$N_L \geq 2$. 

In our scheme, events $A$ and $B$ arise from cascade decays of 
$\widetilde{q}_L^{(\dagger)} \widetilde{q}_L$. In a run of 110 pb$^{-1}$ the
expected number of events with $N_L \geq 2$ is ($\sigma_T / 0.05$ pb)
($\epsilon /0.25$), where $\epsilon$ is the detection efficiency
$\approx 0.25$ and $\sigma_T$
is the total 
$\widetilde{q}_L^\dagger \widetilde{q}_L+
 \widetilde{q}_L \widetilde{q}_L+
 \widetilde{q}_L^\dagger \widetilde{q}_L^\dagger$ production cross
section. There are two contributions to $\sigma_T$ which may be important:
direct $\widetilde{q}_L^\dagger \widetilde{q}_L$ production, and
$\widetilde{q}_L \widetilde{g}$, $\widetilde{q}_L^\dagger \widetilde{g}$ 
production followed
by $\widetilde{g} \rightarrow \widetilde{q}_L^\dagger q, \widetilde{q}_L
q^\dagger$.
The relative importance of these two contributions depends on
$m_{\widetilde{g}}$ and $m_{\widetilde{q}_R}$, which we have not
determined.
For example, with  $m_{\widetilde{g}} = 330$ GeV and  
$m_{\widetilde{q}_R} = m_{\widetilde{q}_L} = 310$ 
GeV, the direct production contributes 0.03 pb to $\sigma_T$, while
squark-gluino production contributes $0.05$ pb to $\sigma_T$.
For these parameters, a further production rate, $\sigma B$, for
dilepton events of 0.04 pb arises from 
$\widetilde{q}_L^{(\dagger)} \widetilde{q}_R^{(\dagger)}$
production, giving a total expectation of about 2.5 events with
$N_L\geq 2$.  If $\widetilde\tau_L$ is degenerate with 
$\widetilde e_L$ and $\widetilde\mu_L$, this would be depleted by a
factor of about 2.

\begin{figure} 
\psfig{figure=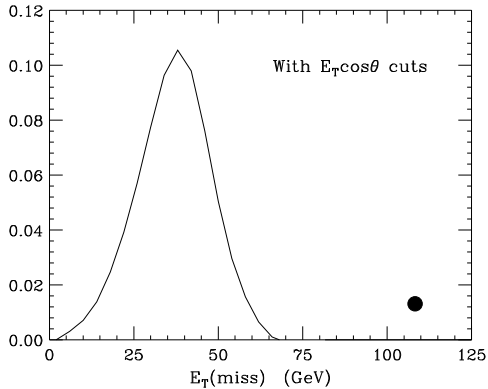,width=2.5in} 
\caption{
Missing $E_T$ for top quark events with cuts allowed by event B
(which is shown as the solid dot): 
$E_T^{\ell_1}\cos\theta(\ell_1-\!\not\!\!E_T) > 100$ GeV and 
$E_T^{\ell_2}\cos\theta(\ell_2-\!\not\!\!E_T) > 40$ GeV.
\label{fig:dist2}}
\end{figure}

Perhaps the most notable result of our analysis is that with only
three candidate events from a hadron collider, we are able to roughly 
estimate the masses of six supersymmetric particles (and the
gaugino/Higgsino content of the $\widetilde\chi'$ states at 260 GeV).  
Clearly more data are needed to refine these estimates and to establish the
particular scenario we have described.

If our scenario is correct, we also anticipate the observation of    
events with large missing $E_T$ and 0, 1, 2, 3, and (very
rarely) 4 leptons 
(though some may have significant backgrounds).
These 1-lepton events may have only two jets  
and hence would not be in the top quark sample. 

Additional signatures may
also be present, depending on the values of $m_{\widetilde{q}_R}$ and 
$m_{\widetilde{g}}$. The production of $\widetilde{g}\widetilde{q}$ contributes
equally to same-sign \cite{SS} and opposite-sign dileptons
($\widetilde{q}^{(\dagger)} \widetilde{q}$ production can also lead to
same-sign events).
When right-handed squarks are
produced, they decay directly to the LSP: $\widetilde{q}_R \rightarrow q_R
\widetilde{\chi}^0_1$, so that several new signals are possible. For example,
with $m_{\widetilde{g}} = 330$ GeV and $m_{\widetilde{q}_R} = 
m_{\widetilde{q}_L} = 310$ GeV, we find a production rate,
$\sigma B$, for ($jj \not
\!\!E_T,jj\ell \not\!\!E_T$) events of (0.13, 0.17) pb.
The standard model backgrounds for these $N_L = 0,1$
events are larger than for the case of $N_L = 2$. However, the signal events
are prominent: the $jj \not\!\!E_T$ events have $E_T^j \sim50-230$ GeV and
$\not\!\!E_T \sim 50-280$ GeV.

%
The reach in squark mass in this scenario exceeds that of several
previous analyses, because the signal can be kinematically
distinguished from the
$\bar{t} t$ background. 
The superpartner masses of our scheme
are so high that no supersymmetric particle would be found at LEP2,
and a 500 GeV NLC would not find all of these particles. If this
turns out to be the first evidence for supersymmetry, the
confirmation will come in the next Tevatron run which may obtain
10-20 times as many events. It may also be possible to identify a few events
with large $\not\!\!E_T$ and 0, 1, 2 same-sign, or three isolated leptons in
the present data.

\section*{Acknowledgments} 
We thank members of the CDF and D0 collaborations for useful conversations.
This work was supported in part by the Director, Office of 
Energy Research, Office of High Energy and Nuclear Physics, Division of 
High Energy Physics of the U.S. Department of Energy under Contract 
DE-AC03-76SF00098 and in part by the National Science Foundation under 
grants PHY-93-20551 and PHY-95-14797.


\begin{thebibliography}{99}
\bibitem{conf} These events have been shown at recent conferences, such as
  SUSY96, and are included in the reported top quark dilepton sample.
\end{thebibliography}
\end{document}